\documentclass[11pt,a4paper]{article}
\usepackage{jheppub,amsmath,  amssymb,slashed,url,bm,textgreek,upgreek}
\usepackage{graphicx}
\usepackage{epstopdf}
\def\t{{ \sf t}} 

\def\tt{{\mathfrak t}}

\def\a{a}

\def\sign{{\mathrm{sign}}}

\def\be{\begin{equation}}
\def\ee{\end{equation}}

\def\a{{\sf a}}

\def\Re{{\mathrm{Re}}}
\def\Im{{\mathrm{Im}}}

\def\h{\widehat}

\def\A{{\mathcal A}}
\def\d{{\mathrm d}}

\def\b{\overline}
\def\R{{\mathbb R}}
\def\C{{\mathbb C}}
\def\U{{\mathcal U}}

\def\[{\bigl [}

\def\]{\bigr ]}

\def\t{\widetilde }
\def\h{\widehat}

\def\B{{\mathcal B}}

\def\i{{\mathrm i}}

\def\H{{\mathcal H}}

\def\U{{\mathcal U}}

\def\E{{\mathcal E}}
\def\add{{\mathrm{add}}}
\def\WF{{\mathrm{WF}}}
\def\la{\langle}
\def\ra{\rangle}
\def\b{{\sf b}}

\font\teneurm=eurm10 \font\seveneurm=eurm7  \font\fiveeurm=eurm5
\newfam\eurmfam
\textfont\eurmfam=\teneurm \scriptfont\eurmfam=\seveneurm
\scriptscriptfont\eurmfam=\fiveeurm

\font\teneusm=eusm10 \font\seveneusm=eusm7 \font\fiveeusm=eusm5
\newfam\eusmfam
\textfont\eusmfam=\teneusm \scriptfont\eusmfam=\seveneusm
\scriptscriptfont\eusmfam=\fiveeusm

\font\tencmmib=cmmib10 \skewchar\tencmmib='177
\font\sevencmmib=cmmib7 \skewchar\sevencmmib='177
\font\fivecmmib=cmmib5 \skewchar\fivecmmib='177
\newfam\cmmibfam
\textfont\cmmibfam=\tencmmib \scriptfont\cmmibfam=\sevencmmib
\scriptscriptfont\cmmibfam=\fivecmmib

\title{The Timelike Tube Theorem in Curved Spacetime}

\author{Alexander Strohmaier$^1$ and Edward Witten$^2$}
\affiliation{$^1$ School of Mathematics,  University of Leeds,  Leeds, Yorkshire, LS2 9JT,UK \\
$^2$ School of Natural Sciences, Institute for Advanced Study,  1 Einstein Drive, Princeton, NJ 08540 USA}
\abstract{The timelike tube theorem asserts that in quantum field theory without gravity, the algebra of observables in an open set $\U$ is the same
as the corresponding algebra of observables in its ``timelike envelope'' $\E(\U)$, which is an open set that is in general larger. 
The theorem was originally proved in the 1960's  by Borchers and Araki for quantum fields in Minkowski space.   Here we sketch the proof of a version
of the theorem for quantum fields in a general real analytic spacetime.   Details have appeared elsewhere.
}

\begin{document}\maketitle

\section{Introduction}\label{intro} 

In ordinary quantum field theory without gravity, one can associate an algebra of observables $\A(\U)$ to any open set $\U$ in spacetime.\footnote{In the present article,
spacetime is always assumed to be globally hyperbolic.}   However, there are two
principles that assert, under certain conditions, that the algebra of some given open set is the same as the algebra of some larger open set.

The first and most familiar is relativistic causality.   Let $D(\U)$ be the domain of dependence of the open set $\U$.  Classically, one would say that fields in $D(\U)$ 
are uniquely determined by fields in $\U$.  The quantum analog of this assertion is the statement that $\A(\U)=\A(D(\U))$.  

\begin{figure}
 \begin{center}
   \includegraphics[width=4in]{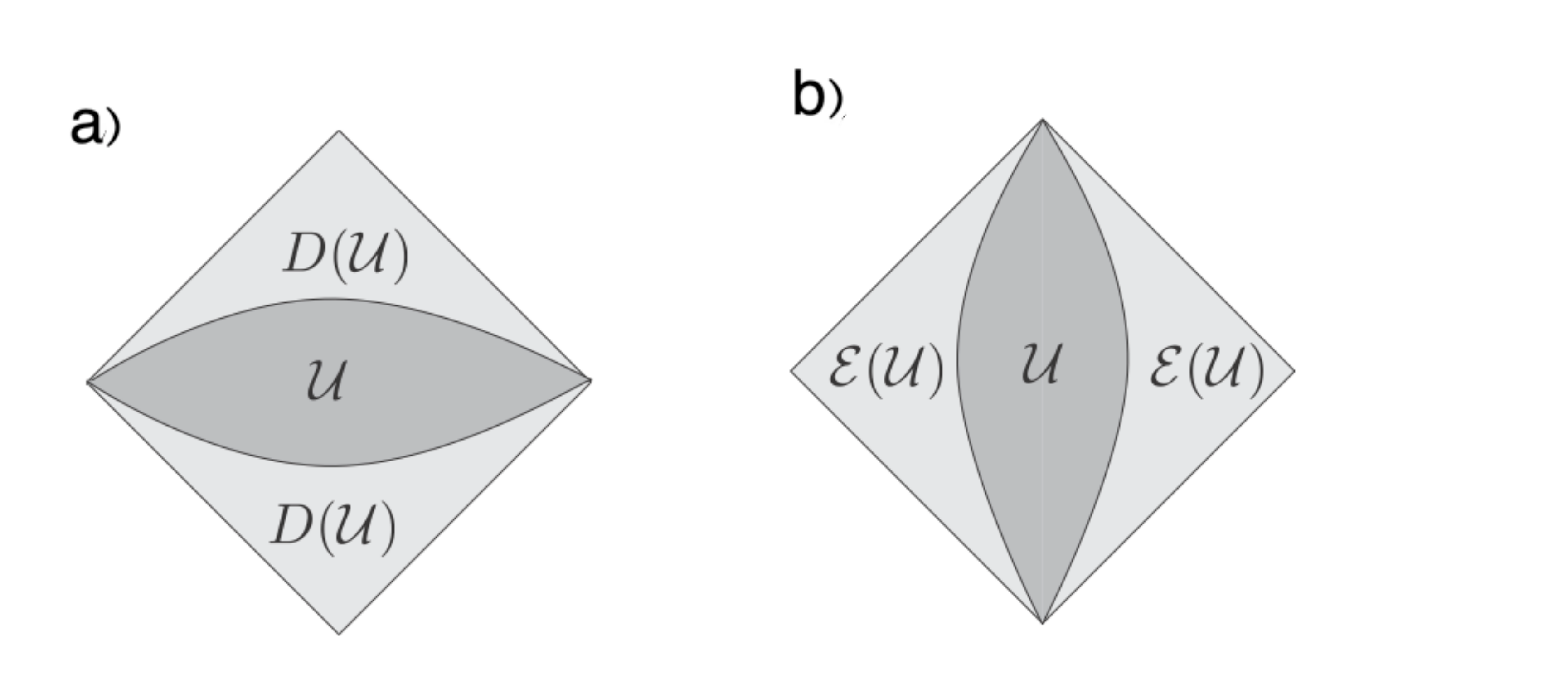}
 \end{center}
\caption{\footnotesize  In these figures, time runs vertically.   Illustrated in (a) is the extension  in a timelike direction
from $\U$ to $D(\U)$ and in (b) the extension in a spacelike direction from $\U$ to $\E(\U)$.     \label{one}}
\end{figure}

 However, in a real analytic spacetime,
the extension from $\U$ to $D(\U)$  has a remarkable ``dual'' version, given by the timelike tube theorem \cite{borchers,araki,stroh,strohwitten}.  The statement of
this theorem involves 
 the {\it timelike envelope} $\E(\U)$ of an open set $\U$, which is defined to consist of all points that can be reached by starting with a timelike curve 
$\gamma\subset \U$ and deforming it through a
family of timelike curves, keeping its endpoints fixed.    The timelike tube theorem asserts that, in a real 
analytic\footnote{By contrast,  causality --
the extension from $\U$ to $D(\U)$ -- does not depend on  real analyticity.}   spacetime $M$,  $\A(U)=\A(\E(\U))$.     For examples of the relation of $\U$ to $D(\U)$ or $\E(\U)$, see fig.
\ref{one}.

The timelike tube theorem can be viewed as a quantum version of the Holmgren uniqueness theorem for partial differential equations.  (For an accessible
account of this theorem, see \cite{Smoller}.) 
According to this theorem, in a real analytic spacetime, if a solution is given in $\U$ of any standard relativistic wave equation, 
such as Maxwell's equations, then the extension of
this solution over $\E(\U)$ is unique, if it exists.     The quantum analog of this statement is that operators in $\E(\U)$ are  determined
by  operators in $\U$ in the sense that $\A(\E(\U))=\A(\U)$.  For some further qualitative discussion, see \cite{WittenLecture}.

  The timelike tube theorem was originally proved by Borchers \cite{borchers} and Araki \cite{araki} for quantum fields in 
  Minkowski space.   The proof by Borchers relied on fairly subtle
  properties of holomorphic functions of several complex variables.   Araki's proof relied on Holmgren uniqueness, more specifically on the fact that a solution of the 
  massless
  scalar wave equation that vanishes in an open
  set $\U$ also vanishes in $\E(\U)$.   (The theory under study was not assumed to be free or weakly coupled, but nonetheless the theorem was deduced
  from properties of a linear wave equation.)      The  timelike tube theorem for free field theories 
  in a real analytic curved spacetime\footnote{The hypothesis required was  somewhat weaker
than real analyticity.   It was sufficient to know that, in some coordinate system, 
the spacetime metric depends on the time in a real analytic fashion.}
 was proved by one of us in \cite{stroh}.

 In \cite{strohwitten}, we have proved a version of the theorem for not necessarily free theories in a  real analytic curved spacetime.   
  The purpose of the present article is to give an informal explanation of this proof.   In section \ref{formulation}, we explain a 
  convenient formulation of the timelike tube theorem
  that has been used in all work on the subject.   In section \ref{microlocal}, we explain the tools in microlocal analysis that are 
  needed in the proof.   The results in microlocal
  analysis that we require imply the facts that were used in Araki's proof, but are more general.   In section \ref{proof}, we 
  explain the proof.   As we will see, the proof of the
  timelike tube theorem in curved spacetime is rather short, once the appropriate tools are available and once one 
  understands what one should be assuming about quantum fields
  in curved spacetime.   In particular, the proof requires the concept of an ``analytic state,'' which is roughly a state from which it is exponentially unlikely
  to extract an asymptotically large amount of energy.   The notion of an analytic state also suffices for the proof in  \cite{SVW} of  a version of the
  Reeh-Schlieder theorem in curved spacetime (though a stronger condition was actually assumed in that paper).
  The Reeh-Schlieder theorem  is the basic result governing entanglement in quantum field theory; we sketch its proof in curved spacetime
  in  section \ref{rs}. In section
  \ref{analytic}, we elaborate on the notion of an analytic state and explain why one would expect that, in a real analytic spacetime, the Hilbert space of a theory
  has a dense set of analytic states.    
  
  In this introduction, we have elided an important point.   As will become clear when we explain the proof,   
the algebra of an open set  that enters in the
  timelike tube theorem is the algebra generated by local fields.   In  contemporary literature, 
  this is sometimes called the additive algebra $\A_\add(\U)$  \cite{casini1,casini2}, but in the present article we simplify the notation by writing simply
  $\A$ rather than $\A_\add$.   In some quantum field
  theories, the full algebra $\A(\U)$ of observables in a general open set $\U$ is slightly larger
  than $\A_\add(\U)$, because of operators such 
  as topologically nontrivial Wilson loop operators that can be defined in the region $\U$ but cannot be constructed from local fields.    In general,
$\A(\U)$  can be constructed from $\A_\add(\U)$ given a knowledge of topological properties of the theory and of
  the region $\U$.   
  It is believed that the distinction between $\A_\add(\U)$ and $\A(\U)$ is absent  in theories that arise as long distance approximations to
   ultraviolet complete theories of quantum gravity. 
 See  \cite{WittenLecture} for further discussion and references.
   
  \section{A Useful Reformulation}\label{formulation}
  
What are the operators in an open set $\U$?  Most obvious are the operators  that are directly constructed from local fields.
  For example, if $\phi$ is a real scalar field in the theory under consideration, and $f$ is a smooth function with support in $\U$, we can consider the operator
  $\phi_f=\int \d\mu f(x) \phi(x)$ (where $\d\mu$ is the Riemannian measure of the spacetime $M$).   This makes sense as a densely-defined unbounded operator.
  Bounded functions of $\phi_f$, such as $\exp(\i \phi_f)$ or in general $F(\phi_f)$, where $F$ is any bounded function of a real variable, are bounded operators.
  To define an algebra of operators, one really wants to consider bounded operators, because bounded operators are defined on all of Hilbert space, and can be
  straightforwardly added or multiplied to define an algebra.
  
Let  $\A_0(\U)$ be the algebra of polynomial functions of operators of the form $F(\phi_f)$.   We might think of $\A_0(\U)$ as an algebra of
  simple operators in $\U$.    However, a simple statement like causality $\A(\U)=\A(D(\U))$ or the timelike tube theorem $\A(\U)=\A(\E(\U))$ will not be true
  if by $\A(\U)$ we mean an algebra of simple operators such as those in $\A_0(\U)$.   The reason is that a simple operator in $D(\U)$ or $\E(\U)$ might
  be equivalent to a more complicated operator in $\U$.   For causality or the timelike tube theorem, we need to consider all operators (or at least all
  ``topologically trivial''  operators, in a sense explained in the last paragraph of the introduction) in the region $\U$.
    One approach to getting all operators is to ``complete'' $\A_0(\U)$ by taking limits.   The relevant notion of limit is defined by convergence of
  matrix elements; if $\a$ is a bounded operator, we say that a sequence $\a_1,\a_2,\cdots\in\A_0(\U)$ converges to $\a$
   if $\lim_{i\to\infty}\la \chi|\a_i|\psi\ra=\la\chi|\a|\psi\ra$
  for all states $\psi,\chi$ in the Hilbert space $\H$ of the theory.    By adjoining such limits, we complete $\A_0(\U)$ to an algebra $\A(\U)$  
  that (assuming we consider all possible local fields in this construction) is in fact the algebra to which the timelike tube theorem applies.   
  $\A(\U)$ is a von Neumann algebra, which just means an algebra of bounded operators that is closed under limits and under taking 
  adjoints.\footnote{See \cite{Sorce} for
  a thorough introduction to von Neumann algebras for physicists.}
  
  However, a slightly different description of $\A(\U)$ is more useful.   In general, if $\B$ is any collection of operators,
  one defines its commutant $\B'$ to consist of all (bounded) operators that commute with $\B$.   $\B'$ is always an algebra, 
  because the sum or product of two operators that commute with $\B$ also commutes with $\B$; and in fact (assuming that $\B$ is closed under taking
  adjoints), it is a von Neumann algebra,
  since a limit of operators that commute with $\B$ also commutes with $\B$.   We can iterate this construction and define $\B''=(\B')'$, the commutant of
  $\B'$.   One has always $\B\subset \B''$ ($\B'$ was defined to commute with $\B$, while $\B''$ consists of all operators that commute with $\B'$,
  so $\B\subset \B''$).   $\B''$ is always a von Neumann algebra, since the commutant of anything is a von Neumann algebra, as already noted.
  In fact, $\B''$ is the smallest von Neumann algebra containing $\B$.  A special case of this statement is an important result of von Neumann: if $\B$
  is already a von Neumann algebra then $\B''=\B$.  In particular, since $\B'$ is always a von Neumann algebra, one has always $\B'=\B'''$.
  
  Going back to the algebra $\A_0(\U)$ of simple operators in $\U$, its completion $\A(\U)$ by taking limits is the smallest von Neumann algebra containing
  $\A_0(\U)$ and therefore is the same as $\A_0(\U)''$.  Likewise $\A(\E(\U))=\A_0(\E(\U))''$.    Hence the timelike tube theorem $\A(\U)=\A(\E(\U))$ is equivalent to 
  $\A_0(\U)''= \A_0(\E(\U))''$.    But this statement follows from $\A_0(\U)'=\A_0(\E(\U))'$, by taking commutants.    The converse of this is also true;
  if $\A_0(\U)''=\A_0(\E(\U))''$, then, taking commutants again, $\A_0(\U)'''=\A_0(\E(\U))'''$.   But as $\B'=\B'''$ for all $\B$, this implies that $\A_0(\U)'=\A_0(\E(\U))'$.
  
  The upshot is that the timelike tube theorem is completely equivalent to the statement that $\A_0(\U)'=\A_0(\E(\U))'$.   In other words, the timelike tube theorem
  is equivalent to the statement that an operator $\b$ commutes
  with all operators in $\A_0(\U)$ if and only if it commutes with all operators in $\A_0(\E(\U))$.   Recalling that $\A_0(\U)$ and $\A_0(\E(\U))$ are generated
  by local fields, an equivalent statement is that if an operator $\b$ commutes with all local operators $\phi(x)$ with $x\in\U$ then it commutes with all local
  operators $\phi(x)$ with $x\in \E(\U)$.
  
 In fact, here it suffices to consider matrix elements of this statement among a dense set of states.   So finally we arrive at the formulation of the timelike
 tube theorem that is most useful in practice.   The timelike tube theorem is true if there is a dense set $S$ of states
with the property that if $\b$ is an operator such that
\be\label{matemt}\la \chi|[\phi(x),\b]|\psi\ra=0 \ee
for all $\chi,\psi\in S$ and all local operators $\phi(x)$ with $x\in\U$, then in fact the same matrix elements vanish for all $x$ in the possibly
larger open set $\E(\U)$.  

In other words, if the timelike tube theorem is 
false, this means that there is some matrix element (\ref{matemt}) that vanishes identically in the open set $\U$, but not in the larger open set $\E(\U))$.
In a real analytic spacetime, if this occurs, that represents a failure of real analyticity.  To prove the timelike tube theorem, we will 
need to know something about the extent to which matrix elements or correlation functions of local operators can fail to be real analytic.
  
  \section{The Wavefront Set}\label{microlocal}
  
  Let $\phi(x)$ be a function, or more generally a distribution, on $\R^n$.  Suppose first that we are interested in whether  $\phi$ is smooth,
  meaning that it can be differentiated any number of times.   Assuming that $\phi$ and its derivatives
  vanish sufficiently rapidly at infinity, there is a well-known criterion for smoothness: $\phi$ is smooth if and only if its Fourier 
  transform $\t\phi(p)$ vanishes faster than any power of $p$ for $p\to\infty$.
  In one direction, this follows from the formula
  \be\label{fourtr} \t\phi(p)=\int_{\R^n}\d^n x \,e^{-\i p\cdot x}\phi(x) = \int_{\R^n}\d^n x \,e^{-\i p\cdot x} \left( -\frac{\i}{p^2} p\cdot\frac{\partial}{\partial x}\right)^k
  \phi(x), \ee
  which holds for arbitrary  positive integer $k$ (if there is no problem in integrating by parts) and
  shows that if all derivatives of $\phi$ exist and vanish fast enough at infinity, then $\t\phi(p)$ vanishes faster than any power of $p$.   In the
  opposite direction, one has the Fourier inversion formula
  \be\label{invform}\phi(x)=\int\frac{\d^n p}{(2\pi)^n} e^{\i p\cdot x}\t\phi(p). \ee
  If $\t\phi(p)$ vanishes faster than any power of $p$, this formula can be differentiated with respect to $x$ any number of times, showing that $\phi$ is smooth.   
  
  As a criterion for smoothness, this has two drawbacks.   First, $\phi$ might not behave well enough at infinity to make this discussion applicable; second,
  we might be interested in the smoothness of a function or distribution $\phi$ on a general manifold $M$, on which there is no natural notion of a Fourier
  transform.   There is a simple way to circumvent this difficulty.   Let $F$ be any smooth function supported in a ball around a point  $x_0$
 where we want to probe for the smoothness of $\phi(x)$ and equal to 1 in a neighborhood of $x_0$.  
 Then $\phi$ is smooth at $x=x_0$ if and only if $F\phi$ is smooth at $x=x_0$.   As $F\phi$ has compact support, there is no problem in defining
 its Fourier transform or in integrating by parts,
 and moreover since a small ball in any manifold $M$
 can be embedded in $\R^n$, the discussion is applicable for any $M$.   Thus we learn in general that a function or
 distribution $\phi$ is smooth at $x=x_0$ if and only if there is a cutoff function $F$ that is 1 in a neighborhood of $x_0$ and such that the Fourier transform
 $\t{F\phi}(p)$ vanishes at large $p$ faster than any power of $p$.   
 
 If $\phi$ is not smooth at $x=x_0$, we can get more refined information by asking in which directions in momentum space $\t{F\phi}(p)$ fails to decay faster
 than any power of $p$.   A convenient formulation is to introduce a positive parameter $h$ and, keeping $p\not=0$ fixed, to consider the behavior 
 of $\t{F\phi}(p/h)$ for $h\to 0$.   
 Then $\phi$ is smooth at $x$ if and only if, for some cutoff function $F$ that is 1 near $x$, and for all $p\not=0$, there is a bound
 \be\label{ftbound}|\t{F\phi}(p/h)|<C_N h^N \ee
 for any $N>0$, with an $N$-dependent constant $C_N$.    
We say that a nonzero momentum vector $p$ is a wavefront vector of $\phi$ at $x$ if such a bound does not hold, regardless of the choice of $F$.     The set
of wavefront vectors is defined to be a closed set, so $p$ is also defined to be a wavefront vector at $x$ if there are vectors $p_i$ arbitrarily close to $p$ such that
a bound (\ref{ftbound}) does not hold.    
The wavefront vectors at $x$ capture information about the directions in which $\phi$ fails to be smooth at $x$.
The {\it wavefront set} $\WF(\phi)$ of
 a function or distribution $\phi$ consists
 of all phase space points $(x,p)$ such that $p$ is a wavefront vector at $x$.   $\WF(\phi)$ does not depend on the coordinate system used to define the
 Fourier transform, and   is a well-defined subset of the cotangent bundle $T^*M$.   Note that $\WF(\phi)$ is always invariant under rescaling of $p$ and that by definition $p$ is required to be nonzero.

Suppose that we are interested in assessing the real analyticity, rather than smoothness, of $\phi$.   A function or distribution $\phi$ on $\R^n$ is real
analytic if it can be analytically continued to a neighborhood of $\R^n\subset \C^n$.     As in the case of  smoothness, if $\phi$ behaves
sufficiently well at infinity, there is a simple criterion for real analyticity in terms of
the Fourier transform $\t\phi(p)$:   $\phi$ is analytic in a neighborhood $|{\mathrm{Im}}\,x|<\epsilon$ of $\R^n$ if and only if $\t\phi(p)$ vanishes
exponentially for $p\to \infty$.   In one direction, if $\t\phi(p)$ decays exponentially for large $p$, the inversion formula (\ref{invform}) remains convergent
if $x$ has a small imaginary part and shows the claimed real analyticity of $\phi$.  In the other direction, if $\phi$ is holomorphic for $|\Im\,x|<\epsilon$ and decays
fast enough for $\Re\,x\to\infty$, then a simple
shift of the contour in eqn. (\ref{fourtr}) by $x\to x-\i \alpha p/|p|$ (for small positive $\alpha$)  shows that $\t\phi(p)$ vanishes exponentially for large $p$.  

As in the discussion of smoothness, this criterion for real analyticity has two drawbacks.   First, we would like a local criterion for real analyticity of $\phi(x)$
irrespective of how $\phi$ behaves for $x\to\infty$.   Second, we want a criterion that makes sense on a general manifold $M$.   Here actually
we should pause for a moment to explain what it means for a spacetime $M$ to be real analytic and for a function on $M$ to be real analytic.
A real analytic structure on $M$ can be defined by specifying an embedding of $M$ as a real subspace of a complex manifold\footnote{No global properties of $M_\C$
are assumed; in general $M_\C$ might develop very severe singularities outside a small neighborhood of $M$.}  $M_\C$ of the same
dimension (thus, the embedding should be such that there is
 an antiholomorphic symmetry $\lambda:M_\C\to M_\C$ with $\lambda^2=1$ and $M$ as the fixed point set of $\lambda$).   
Then functions on $M$ are called real analytic if they can be analytically
continued to holomorphic functions on a neighborhood of $M\subset M_\C$.   A  Riemannian (or pseudo-Riemannian) manifold $M$ is called real analytic 
if in a real analytic coordinate system on $M$, the components of the metric tensor are real analytic.

In contrast to the discussion of smoothness, we cannot get a criterion for real analyticity of $\phi$ by replacing $\phi$ with $F\phi$, for a compactly
supported cutoff function $F$.   Since a compactly supported $F$ is never real analytic, it will never happen that $\t{F\phi}(p)$ decays exponentially for $p\to\infty$.
However, there is a simple cure for this: instead of the Fourier transform, we have to use the Fourier-Bros-Iagolnitzer (FBI) transform.   While 
the Fourier transform of a function $\phi(x)$ is a function on momentum space, the FBI transform of $\phi(x)$ is a function on phase space.  The FBI transform
also depends on a positive parameter $h$, usually taken to be asymptotically small.   The FBI transform $T_h \phi$ of a function $\phi$ is defined 
by\footnote{A multiplicative factor (depending only on $h$)
 is often included in this formula to ensure that $T_h$ is an isometry from functions on $\R^n$ to functions on 
phase space.   We omit this factor, which would play no role in our discussion.}
\be\label{fbi}(T_h \phi)(x,p)=\int \d^n y \,e^{-(x-y)^2/2h -\i p\cdot y/h} \phi(y).    \ee 
In terms of the FBI transform, there is a simple criterion for real analyticity of $\phi$:  $\phi$ is real analytic in a neighborhood of a point $x$
if and only if, for some cutoff function $F$ that is 1 in a neighborhood of $x$ and any nonzero $p$,
 $(T_h(F\phi))(x,p)$  vanishes exponentially as $h\to 0$.   This exponential vanishing, to be precise, means that 
 \be\label{nfbi} \bigl|(T_h(F\phi))(x,p)\bigr|< C e^{-\delta/h} \ee
with positive constants $C,\delta$ which can be chosen to depend continuously on $x$ and $p$.    In one direction, if $\phi$ is real analytic at $x$, then a bound as in (\ref{nfbi}) is easily proved by deforming the integration
cycle in eqn. (\ref{fbi}) by giving $y$ a small imaginary part near $x$.   In the opposite direction, to show that a bound (\ref{nfbi}) implies that $\phi$ is
real analytic at $x$ requires a more involved contour deformation argument.  Qualitatively, the reason that such a result is possible
 with the FBI transform and not with the Fourier transform is that, if $F$ is a cutoff function that is 1 near $x$,
then because of the Gaussian factor $e^{-(x-y)^2/2h}$, the contribution to 
\be\label{fbin}(T_h (F\phi))(x,p)=\int \d^n y \,e^{-(x-y)^2/2h -\i p\cdot y/h} F(y)\phi(y).    \ee 
from the region in which $F(y)$  is not real analytic is exponentially small in $h$.  Hence for the FBI transform, unlike the Fourier transform, modulo an exponentially small
error it is not important that the cutoff violates real analyticity.

As in the case of smoothness, we say that $p\not=0$ is an analytic wavefront vector of $\phi$ at $x$ if a bound as in (\ref{nfbi}) does not hold, for any choice of $F$ (and more
generally if that is the case for vectors $p_i$ that are arbitrarily close to $p$).   
The analytic wavefront vectors at $x$ capture information about how $\phi$ fails to be real analytic at $x$.   
The analytic wavefront set $\WF_a(\phi)$ consists of all phase space points $(x,p)$ such that $p$ is an analytic wavefront vector at $x$.  As in the smooth
case, $\WF_a(\phi)$ does not depend on the coordinate system used to define the FBI transform, and is a well-defined subset of $T^*M$, invariant
under rescaling of $p$.

   \begin{figure}
 \begin{center}
   \includegraphics[width=3in]{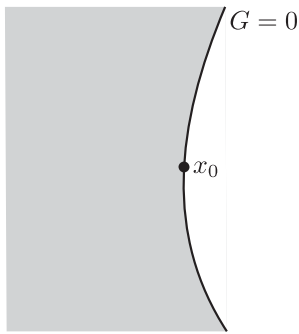}
 \end{center}
\caption{\footnotesize  A distribution $\phi$ vanishes to the left of the hypersurface $G=0$ and not to the right.   $\phi$ does not vanish in any neighborhood
of the point $x_0$.  \label{two}}
\end{figure} 
For our purposes, the most important property of the analytic wavefront set is the following.   Suppose that a function or distribution $\phi$ vanishes identically 
on one side of a hypersurface $H$ and not on the other side.   For example, if $H$ is defined locally by the vanishing of a function $G$, we can assume
that $\phi$ vanishes identically for $G>0$ but not for $G<0$.  Let $x_0$ be a point in $H$ such that $\phi$ does not vanish identically in any neighborhood
of $x_0$ (fig. \ref{two}).   The differential $\d G$ of $G$ can be interpreted as a momentum vector along $G$.   Then 
\be\label{bothsides} (x_0,\pm \d G)\in \WF_a(\phi). \ee
   Since $\WF_a(\phi)$ is
invariant under scaling of $p$, it actually contains points $(x_0,t \,\d G) $ for any nonzero real $t$.
The assertion (\ref{bothsides})  follows from Theorem 8.5.6$'$ of \cite{Hormander}, but we will just illustrate the idea with some representative examples.

A basic example, in one dimension, of a function that  vanishes on one side of a hypersurface and not on the other is 
\be\label{thetafun} \uptheta(x)=\begin{cases} 1 & x\geq 0 \cr 0 & x<0 .\end{cases}\ee
   Here as in the following examples, the behavior of $\uptheta(x)$ for large $|x|$ is mild enough so that
 in analyzing the FBI transform, it is not important to include a cutoff function.
One has
\be\label{fbitheta} (T_h\uptheta)(x,p)=\int_0^\infty \d y \, e^{-(x-y)^2/h-\i p y/h}. \ee 
For any $x\not=0$ and any nonzero $p$, this integral is exponentially small for $h\to 0$.   However, at $x=0$ and $p\not=0$, the integral
is not exponentially small.   In fact, it is asymptotic to $h/\i p$ for $h\to 0$.   Thus $\WF_a(\uptheta)$  consists of phase space points
$(x,p)=(0,p)$ with any nonzero $p$.   There are no wavefront vectors $(x,p)$ with $x\not=0$ because $\uptheta(x)$ is real analytic except at $x=0$,
and there are wavefront vectors $(0,p)$ for either sign of $p$ in keeping with the assertion in eqn. (\ref{bothsides}).

A more challenging example of a function that vanishes on one side of a hypersurface and not the other is a smooth function that vanishes
identically on one side of the hypersurface.   For example, in one dimension, we can consider
\be\label{otherex} \phi(x) = \begin{cases} \exp(-1/x) & x\geq 0 \cr 0 & x<0 .\end{cases}\ee
$T_h\phi(x,p)$ vanishes exponentially for $x\not=0$ because of real
analyticity.  At $x=0$, one has
\be\label{subtlercase} T_h \phi(0,p) = \int_0^\infty \d y \exp(-y^2/2h-\i p y/h -1/y). \ee 
The exponent $f(y)= -y^2/2h-\i p y/h -1/y$ has three critical points in the complex plane, which for small $h$ are at approximately $y=-\i p$ and $y=\pm(\i p/h)^{1/2}$.
Associated to each critical point $q$ is a steepest descent contour $\gamma_q$, with the property that for small $h$, 
$\int_{\gamma_q} \d y  \exp(-y^2/2h-\i p y/h -1/y)$  can be approximated by an integral near  $q$.   The integral (\ref{subtlercase})
can be expressed by contour deformation as a linear combination of the three steepest descent integrals, with the conclusion that 
\be\label{asymptf} (T_h\phi)(0,p)\overset{h\to 0}{\sim}\exp(\i \pi/4) h^{3/2} \sqrt{\pi} \exp(\i/2)\exp(-2 e^{\i \pi\sign(p)/4}(|p|/h)^{1/2}),\ee  dominated by the value at one of the three critical points.   
Thus, $(T_h\phi)(0,p)$ vanishes for $h\to 0$ faster than any power of $h$, but not exponentially fast.    $\WF_a(\phi)$ therefore
contains the phase space points $(0,p)$ for any nonzero $p$.   This is expected, since $\phi(x)$ is nonzero for $x>0$ and vanishes for $x<0$.   The reader
can verify that by contrast the ordinary or smooth wavefront set $\WF(\phi)$ is empty, in keeping with the fact that $\phi$ is smooth.

We will consider one more example, chosen to illustrate the consequences of holomorphy in a half-plane.   Again in one dimension,
consider the distribution
\be\label{upperhalf} \varphi_+(x)=\frac{1}{x+\i\epsilon} = P\frac{1}{x} -\i\pi \delta(x), \ee
which can be analytically continued from real $x$ into the upper half plane.  Because $\varphi_+(x)$ is real analytic for $x\not=0$, $\WF_a(\varphi_+)$
vanishes away from $x=0$.   At $x=0$, we have
\be\label{fbiphi}T_h\varphi_+(0,p) =\int_{-\infty}^\infty \d y e^{-y^2/2h-\i p y/h} \frac{1}{y+\i\epsilon}. \ee
To attempt to show that $T_h\varphi(0,p)$ vanishes exponentially for $h\to 0$, we deform the integration
contour near $y=0$ to give $y$ an imaginary part.  For $p<0$, to make $e^{-\i p y/h}$ small,
we give $y$ a positive imaginary part.  This means deforming the contour in the upper half plane,
where $\varphi_+$ is holomorphic; we succeed in proving that $T_h\varphi_+(0,p)$ vanishes exponentially for $h\to 0$.   For $p>0$, to make $e^{-\i p y/h}$ small,
 we have to deform
the contour in the lower half plane.   We pick up a contribution from the pole at $y=-\i\epsilon$, and we learn that the integral is equal to $-2\pi\i$ plus
an exponentially small remainder.   Thus $\WF_a(\varphi_+)$ consists of the points $(0,p)$ with $p>0$.   The same applies for any distribution $\phi$ that
can be analytically continued in the upper half plane near $x=x_0$ but is not analytic at $x=x_0$:   $\WF_a(\phi)$ must contain some points
$(x_0,p)$, since $\phi$ is not analytic at $x=x_0$; on the other hand, holomorphy of $\phi$ in the upper half plane near $x=x_0$ implies, by
 contour deformation, that $(x_0,p)\notin\WF_a(\phi)$  if $p<0$.   Hence $(x_0,p)\in\WF_a(\phi)$  precisely
if $p>0$.   Of course, for a distribution such as $\varphi_-=1/(x-\i\epsilon)$ that is  holomorphic in the lower half plane, the analytic wavefront set
contains instead only points with $p<0$.

  \section{Analytic Vectors and the Proof of the Timelike Tube Theorem}\label{proof}
  
  \subsection{Analytic Vectors}\label{anavect}
  
  Now let us go back to quantum field theory.   In Minkowski space, a basic axiom of quantum field theory is that there is a unique state $\Omega$, the vacuum state,
  with minimum energy; all other states have positive energy.    In a general curved spacetime $M$, there is no conserved energy, and also no distinguished state such as $\Omega$.
  But in going to curved spacetime, we cannot just discard the condition that in Minkowski space says that the energy is bounded below.   We need to replace it by something that makes
  sense in general and that in Minkowski space reduces to the usual statement.   
  
  In a generic $M$, there is no notion of energy-momentum, but the notion of energy-momentum makes sense asymptotically at high energies and short distances.  To generalize the usual
  axioms of quantum field theory to curved spacetime, we need to somehow make use of this asymptotic fact to state a condition that, in Minkowski space, will reduce to the usual 
  positivity of the energy.    It has been found that this can be conveniently done in terms of the wavefront set \cite{Radz,BFK,HW}.
  
  The basic idea is the following.   Consider the $n$ point function of a local field $\phi$ in the Minkowski space vacuum:
  \be\label{npoint}\upphi(x_1,x_2,\cdots, x_n)=\la\Omega|\phi(x_1)\phi(x_2)\cdots \phi(x_n)|\Omega\ra.\ee
  This is a function (or really a distribution) on $\R^D\times \R^D\times \cdots \times \R^D=\R^{nD}$.   It is not real analytic.   It has singularities due to the emission and
  absorption of particles that can have arbitrarily large energy.   Accordingly, the analytic wavefront set $\WF_a(\upphi)$ is nonempty.   However, using positivity of energy, one
  can prove the following.   If 
  \be\label{phasespace} (x_1,p_1; x_2,p_2; \cdots ; x_n,p_n)\in \WF_a(\upphi)  \ee
then the rightmost nonzero $p_k$ is future-directed causal (the leftmost nonzero $p_k$ is past-directed causal).   In other words, if $p_n$ is nonzero, then
  it is future-directed causal; if $p_n=0$ but $p_{n-1}\not=0$, then $p_{n-1}$ is future-directed causal; and so on.   The rough idea is that the rightmost operator that is creating a state of
  very high energy, leading to a singularity, has to create a state of null or timelike momentum, since those are the states that the theory has.   Similarly the leftmost operator
  that is annihilating a state of asymptotically high energy has to annihilate a state of null or timelike momentum.   
  
  In curved spacetime, we do not have a distinguished state $\Omega$.  But we can consider the $n$-point function in a state $\Psi$:
    \be\label{npointcurved}\upphi(x_1,x_2,\cdots, x_n)=\la\Psi|\phi(x_1)\phi(x_2)\cdots \phi(x_n)|\Psi\ra.\ee
    $\upphi$ will again not be real analytic; it will have singularities due to the possibility that particles of arbitrarily high energy can be created and annihilated by the operators.   
    In general, these singularities will depend on the state $\Psi$.   If the state $\Psi$ has a nonnegligible (not exponentially small) amplitude to contain a particle of asymptotically
    high energy, then this will contribute to the singularities of $\upphi$.  However, if $\Psi$ has exponentially small amplitude to contain a particle of asymptotically high energy, then
    we expect singularities to come only from particles of asymptotically high energy created by one or more of the operators and annihilated by others.   Then we can expect the
    wavefront set to be independent of $\Psi$ and to satisfy conditions somewhat like those that hold for the vacuum in Minkowski space.
    
  This motivates the following definition: in a real analytic spacetime $M$, the state $\Psi$ is ``analytic'' if its analytic wavefront set $\WF_a(\upphi)$ satisfies conditions similar to what
  hold for the vacuum vector in Minkowski space: if $(x_1,p_1; x_2,p_2;\cdots; x_n,p_n)\in \WF_a(\upphi)$, then the rightmost (leftmost) nonzero $p$ is future-directed (past-directed)
  and causal.   This condition roughly states that the probability to extract an asymptotically high energy from the state is exponentially small.   Analytic wavefront set conditions of this type were introduced in \cite{SVW} and
  used to prove the Reeh-Schlieder property (this argument will be sketched in section \ref{rs}).  
  
  The notion of an analytic state suffices for our purposes here, but it is actually a rather weak condition.  
  Authors who have tried to systematically describe the properties of the wavefront set or the analytic wavefront set of the correlation functions of a quantum field theory in curved spacetime
  have proposed much stronger conditions   \cite{BFK,HW}. 
  
  We will describe in section \ref{analytic} some constructions of analytic states.   For now, we just remark that any state defined by a reasonable Euclidean construction, such as the
  Hartle-Hawking state of a black hole, is expected to be analytic.
  
  The analog of an analytic state in ordinary quantum mechanics is a wavefunction that is real analytic as a function of the particle positions.  By this definition,
  analytic states are dense in Hilbert space in ordinary quantum mechanics, and the same is expected in quantum field theory.
  
Before trying to prove the timelike tube theorem, we note a few further properties of analytic states.   Consider a general matrix element of a local field $\phi$:
\be\label{indef} \Lambda(x)=\la \Psi'|\phi(x) |\Psi\ra.\ee
If the states $\Psi$ and $\Psi'$ are both analytic, then $\Lambda(x)$ is a real analytic function: $\phi(x)$ cannot annihilate particles of asymptotically high energy (since these are
absent in $\Psi$) and cannot create such particles (since they would not be absorbed by $\Psi'$). So $\WF_a(\Lambda)$ is empty and $\Lambda(x)$ is real analytic.
  Suppose instead that $\Psi$ is analytic but $\Psi'$ is a completely general state.
Then  $\phi(x)$ can create high energy particles but cannot annihilate them, so if $(x,p)\in \WF_a(\Lambda)$, $p$ must be future-directed causal.   Similarly, if $\Psi'$ is analytic,
then regardless of $\Psi$,   if $(x,p)\in \WF_a(\Lambda)$, then $p$ must be past-directed causal.    More generally, in the case of an $n$-point matrix element
\be\label{zindef}\Lambda(x_1,x_2,\cdots,x_n) = \la\Psi'|\phi(x_1)\phi(x_2)\cdots \phi(x_n)|\Psi\ra,\ee
suppose that $(x_1,p_1;x_2,p_2;\cdots; x_n,p_n)\in\WF_a(\Lambda)$.   If $\Psi$ is analytic, it follows that the rightmost nonzero $p$ is future-directed timelike; if $\Psi'$ is analytic,
it follows that the leftmost nonzero $p$ is past-directed timelike.

\subsection{The Proof}\label{thep}

Now we are ready to discuss the proof of the timelike tube theorem.
Assuming that a theory has a dense set $S$ of analytic states, as is expected to be true for any reasonable theory, then one can give a very short proof of the timelike tube theorem,
  as follows.   As in eqn. (\ref{matemt}), we have to show that if the operator $\b$ has the property that for any $\chi,\psi\in S$,
  the distribution
  \be\label{vanishdist}\Upsilon(x)=\la\chi| [\phi(x),\b]|\psi\ra = \la\chi|\phi(x)|\b\psi\ra-\la \b^\dagger\chi|\phi(x)|\psi\ra \ee 
    vanishes for $x\in \U$, then actually $\Upsilon(x)$ vanishes for $x\in \E(\U)$.   In order for a phase space point $(x,p)$ to be in $\WF_a(\Upsilon)$, it must be
    in the analytic wavefront set of one of the two terms on the right hand side of eqn. (\ref{vanishdist}).    By virtue of what was explained in the discussion of eqn. (\ref{indef}),
    for $(x,p)$ to be in the analytic wavefront set of $\la\chi|\phi(x)|\b\psi\ra$, $p$ must be past-directed causal; for $(x,p)$ to be in the analytic wavefront set
    of $\la \b^\dagger\chi|\phi(x) |\psi\ra$, $p$ must be future-directed causal.  Either way, $p$ is causal -- timelike or null.

       \begin{figure}
 \begin{center}
   \includegraphics[width=3.5in]{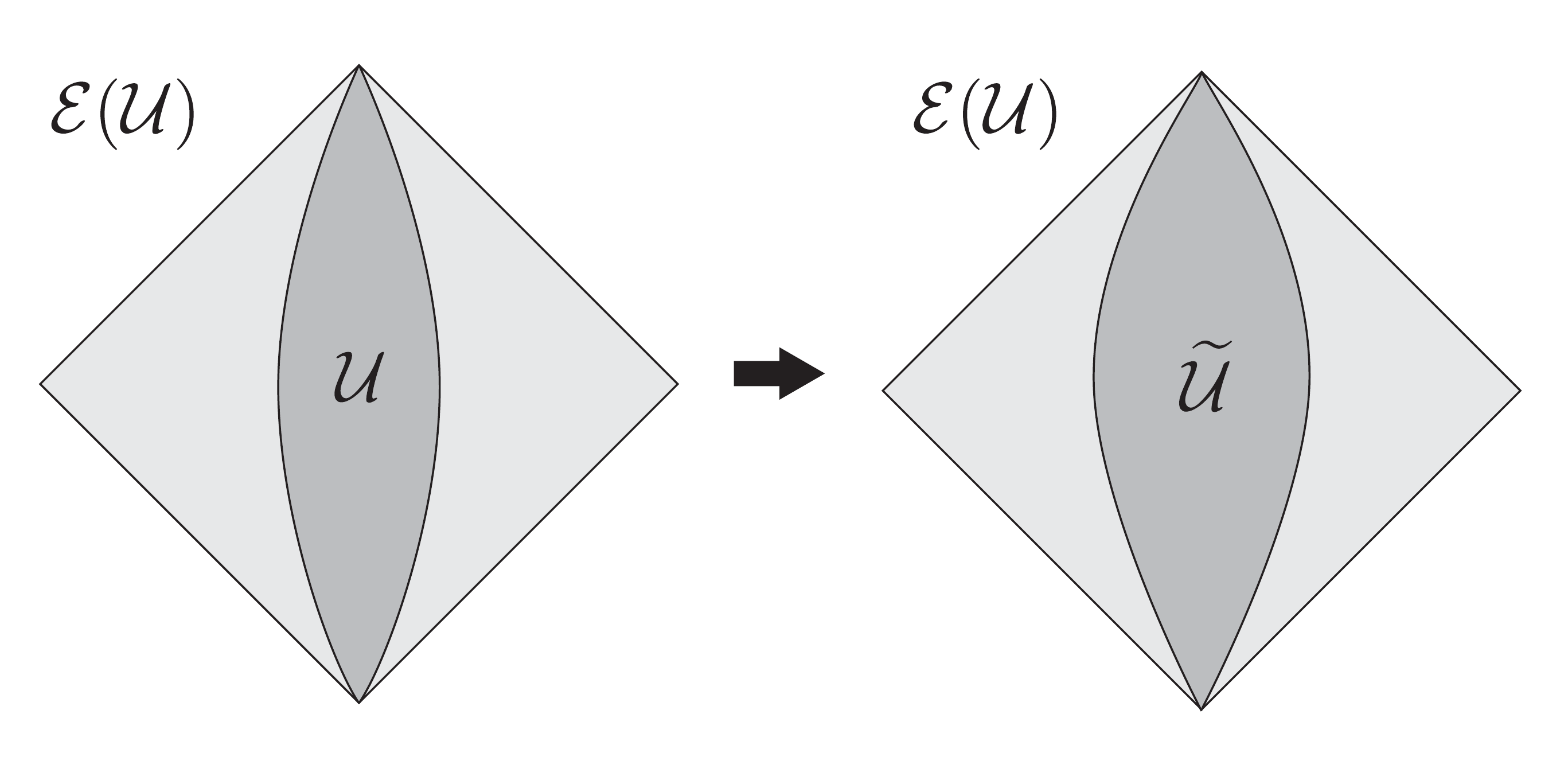}
 \end{center}
\caption{\footnotesize  The hypersurface $\U$ can ``grow'' outward to $\t\U$ in such a way that its normal vector is always everywhere spacelike.   This can continue
until $\U$ expands to fill out the timelike envelope $\E(\U),$ which has null boundaries.    \label{three}}
\end{figure} 
  However, we can ``expand'' $\U$ outward into $\E(\U)$ by gradually growing $\U$ outward in such a way that  the boundary of $\U$ is always a timelike hypersurface $H$, and the normal
  vector $n$ to $H$ is spacelike (fig. \ref{three}).   As in eqn. (\ref{bothsides}), if vanishing of $\Upsilon(x)$ is going to fail along such a hypersurface $H$, then the points $(x,\pm n)$ will be
  in $\WF_a(\Upsilon)$.   This cannot happen, since we just learned that $\WF_a(\Upsilon)$ contains only causal momentum vectors, not spacelike vectors such as $n$.  So $\Upsilon$
  continues to vanish as we expand $\U$ outward until it fills out all of $\E(\U)$.
(We cannot in this way deduce any vanishing beyond $\E(\U)$, since $\E(\U)$ has null boundaries.)  That concludes the proof of the timelike tube theorem.
    
    The argument given in \cite{strohwitten} was slightly different.   Rather than the existence of a dense set of analytic states, what was assumed was the existence 
    of a single analytic state $\Psi$ that also is ``cyclic.''   A vector is called cyclic if the states 
    \be\label{zelbo}\Psi_{x_1,x_2,\cdots, x_n}= \phi(x_1)\phi(x_2)\cdots\phi(x_n)|\Psi\ra\ee are dense
    in the Hilbert space $\H$ of a theory.\footnote{To be more precise, 
    in a theory with superselection sectors because of, for example, the existence of conserved gauge charges,   the $\Psi_{x_1,x_2,\cdots,x_n}$ are
    dense in the vacuum sector of Hilbert space.   For our purposes, since the algebra of observables $\A(\U)$ that appears in the timelike tube theorem is the additive algebra,
    generated by local fields, we work in the vacuum sector.}
  The Wightman axioms assert that the vacuum is a cyclic vector for quantum fields in Minkowski
    space, and it is expected that cyclic vectors exist in quantum field theory in any spacetime.        The existence of a single cyclic vector actually implies that a generic vector is cyclic. 
    The construction of analytic vectors that we will discuss in section \ref{analytic}
    is expected to produce cyclic analytic vectors in any theory.   
    
    Given the existence of a cyclic analytic vector, let $S$ be the dense set of states of the form of eqn. (\ref{zelbo}). The 
    timelike tube theorem can be proved by showing that for $\psi,\chi\in S$, if $\b$ is such that the distribution $\Upsilon(x)$ defined in eqn. (\ref{vanishdist}) vanishes for all $x\in \U$,
    then it vanishes for all $x\in\E(\U)$.   The proof of this in  \cite{strohwitten} was based on ideas that we have already explained, notably  the properties stated earlier for the analytic wavefront set of the distribution
    $\Lambda(x_1,x_2,\cdots, x_n)$ defined in eqn. (\ref{zindef}) and the fact that $\U$ can be expanded outward to fill out $\E(\U)$ in such a way that its normal vector is always
    spacelike.
    
    \subsection{The Reeh-Schlieder Theorem}\label{rs}
        
   What  we have explained in the course of generalizing the timelike tube theorem to curved spacetime also makes it possible to generalize the Reeh-Schlieder theorem \cite{RS}
  to an arbitrary (connected) real analytic spacetime $M$  \cite{SVW}.   As remarked earlier,
  the Reeh-Schlieder theorem is the basic result about entanglement in quantum field theory.       What we have defined as a cyclic
    vector  in eqn. (\ref{zelbo}) is, in more detail, a cyclic vector for the full algebra $\A(M)$ generated by all local fields in $M$.    In other words, the definition of a cyclic vector asserts that
    the states $\Psi_{x_1,x_2,\cdots,x_n}$, with arbitrary $x_i\in M$, are dense.    An equivalent statement is the following: if $\chi$ is any state such that  the inner products
    \be\label{ortho} \Pi(x_1,x_2,\cdots,x_n)=\la\chi|\Psi_{x_1,x_2,\cdots,x_n}\ra =\la\chi|\phi(x_1)\phi(x_2)\cdots \phi(x_n)|\Psi\ra\ee
    vanish for all $x_i\in M$, then $\chi=0$.   Indeed, the existence of a nonzero vector $\chi$ orthogonal to all $\Psi_{x_1,x_2,\cdots,x_n}$ implies that the
    $\Psi_{x_1,x_2,\cdots,x_n}$ are not dense; conversely, if the $\Psi_{x_1,x_2,\cdots,x_n}$ are not dense, then they generate a proper subspace $\H'\subset \H$
    and any vector $\chi$ in the orthocomplement of $\H'$ is orthogonal to all $\Psi_{x_1,x_2,\cdots,x_n}$.
    
    The original Reeh-Schlieder theorem makes the remarkable assertion
     that the vacuum vector of a quantum field theory in Minkowski space
      is cyclic not just for the full field algebra $\A(M)$ but for the algebra $\A(\U)$ of any open set $\U$, no matter
    how small.   In other words, states created by acting on the vacuum with fields in a small open set $\U$ are dense in $\H$.
    The generalization of this theorem to curved spacetime, as formulated in \cite{SVW}, asserts that if an analytic vector $\Psi$ is cyclic for $\A(M)$, then it is cyclic for the  algebra $\A(\U)$ of any open set $\U$.
    In other words, if the states $\Psi_{x_1,x_2,\cdots,x_n}$ are dense, then they remain dense if restricted to $x_i\in \U$.   Equivalently, if  the distribution 
    $\Pi(x_1,x_2,\cdots,x_n)$ on $M^n$ that was defined in eqn. (\ref{ortho}) vanishes
    for all $x_i\in \U$, then it vanishes for all $x_i\in M$.   To prove this, we first consider the constraint on $\WF_a(\Pi)$ that follows from the assumed analyticity of $\Psi$.
    As in the discussion of eqn. (\ref{zindef}), if $(x_1,p_1;x_2,p_2;\cdots; x_n,p_n)\in \WF_a(\Pi)$, then the rightmost nonzero $p_k$ is future-directed and causal.
    This has the following consequence.   Schematically, set $X=(x_1,x_2,\cdots,x_n)$, $P=(p_1,p_2,\cdots,p_n)$.  Then there is no pair $(X,P)$ such that
    $(X,P)\in \WF_a(\Pi)$ and also $(X,-P)\in\WF_a(\Pi)$, since the rightmost nonzero component of any wavefront momentum vector is future-directed.
    
    We can exploit this as follows.   
    Since $M$ is assumed connected, we can continuously enlarge $\U$ until it fills out all of $M$.  (In contrast to the proof of the timelike tube
    theorem, here it does not matter whether $\U$ is expanded outward in timelike or spacelike directions.)   Suppose that the distribution $\Pi$ vanishes identically
    if all points $x_i$ are in  $\U$. In other words, suppose that as a distribution on $M^n$, $\Pi$ vanishes on $\U^n$.
     Does this continue to be true as  $\U$ expands outward?   If $\U$ can be enlarged to $\h\U$ with $\Pi$ still vanishing in $\h\U^n$,
    but ceases to vanish if $\h\U$ is enlarged further, then let $H$ be the boundary of $\h\U^n\subset M^n$.   $\Pi$ is then a distribution on $M^n$ that vanishes identically
    on one side of $H$ and not on the other side.   As in the discussion of eqn. (\ref{bothsides}), it follows that if $H$ is defined locally by an equation $G=0$,
    and $X_0$ is a point in $H$ such that $\Pi$ does not vanish identically
    in any neighborhood of $X_0$,  then the phase space points $(X_0,\pm \d G)$ are both in $\WF_a(\Pi)$.   But we learned in the last paragraph
    that there is no pair $(X,P)$ with $(X,P)$ and $(X,-P)$ both in $\WF_a(\Pi)$.  So vanishing of $\Pi$ is never lost as $\U$ is continuously enlarged.   Thus $\Pi$ is identically zero
    and the Reeh-Schlieder theorem holds.

  \section{Construction of Analytic Vectors}\label{analytic}
  
  In this concluding section, we explain some additional facts to help orient the reader to the notion of an analyic state.
    
  \subsection{Tempered Analytic States}\label{tempered}
  
  In Minkowski space, there is a natural distinguished state, the vacuum vector $\Omega$.   In a general time-dependent spacetime, with no conserved energy, there is
   no natural state and there is not even a small preferred set of natural states.
     In particular, it is not reasonable to expect that, in a general real analytic spacetime $M$, a theory has a unique analytic state, or a distinguished small set of such states.
   If analytic states exist at all, there must be many of them.
   
   This line of thought leads to the following question: given one analytic state, can we construct more?   An obvious idea is that given an analytic state $\Psi$,
    one might be able to make
 additional analytic states by acting on $\Psi$ by a field operator, smeared by a real analytic function.   Thus, if $f$ is a real analytic function on $M$ that vanishes sufficiently
 rapidly at infinity (where here ``vanishing sufficiently rapidly at infinity'' includes a condition on the behavior of $f$ near singularities of $M$, if any), and $\phi_f=\int_M\d \mu f(x)\phi(x)$ is
 a corresponding smeared field,
 then one can hope that $\phi_f\Psi$ will be again an analytic state.  If true, this is useful, because in a real analytic spacetime,
 there are many  real analytic functions that vanish rapidly at infinity.  For example, in Minkowski space with coordinates $t,\vec x$, a simple example of a real analytic function that vanishes
 rapidly at infinity is a Gaussian function $\exp(-t^2-\vec x^2)$. This can be multiplied by any polynomial to produce a very large set of rapidly decaying real analytic functions. In 
 fact,  among smooth functions that vanish rapidly at infinity, the real analytic ones are dense; this is true in any real
 analytic spacetime.
 
 So if acting on an analytic vector with an operator such as $\phi_f$ will give a new analytic vector, then there will be many analytic vectors.
However, it appears difficult to prove this based only on the Wightman axioms of quantum 
 field theory.\footnote{A counterexample at the end of Appendix C of \cite{strohwitten} shows the difficulty in one attempt at a proof,}
 As a substitute, in  \cite{strohwitten}, we defined a notion of a tempered analytic state, which is roughly an analytic state that behaves sufficiently well at infinity.  
  For example, the vacuum state in Minkowski space is tempered analytic. 
  If $\Psi$ is a tempered analytic state and $f_1,\cdots,f_n$ are  real analytic functions that vanish sufficiently rapidly at infinity, then if is possible to prove (Theorem 4.6 in 
  \cite{strohwitten}) that 
the vectors $\phi_{f_1}\phi_{f_2}\cdots \phi_{f_n}\Psi$ are tempered analytic.   
So if a theory has a single tempered analytic state, then it has many such states.   Moreover, if a theory has a single
tempered analytic state that is cyclic for the field algebra, then it has a dense set of tempered analytic states.   For a state to be cyclic for the full field algebra is a mild
condition that is satisfied by a generic state.

If $M$ satisfies certain technical conditions, then in free field theory on $M$, every analytic state is tempered analytic (Proposition 4.9 in \cite{strohwitten}).   
It is conceivable that this is also true in non-free theories that satisfy physically
realistic conditions,  even if the statement does not follow from the Wightman axioms.   Physically realistic theories satisfy conditions (such as the existence of an energy-momentum
tensor) that go beyond the Wightman axioms.

\subsection{Direct Construction Of Analytic States}\label{direct}

In free field theory, there  is a direct rigorous construction of analytic states \cite{GW}.  One actually expects this construction to be applicable in general in realistic quantum field theories.

The construction is easiest to explain for a spacetime $M$ that has a Cauchy hypersurface $S$ that is invariant under a time-reflection symmetry $\varrho$.   We can assume that
$M$ has a real analytic time coordinate $t$ such that $\varrho$ acts by $t\to -t$, leaving fixed $S$.

By definition, a real analytic spacetime $M$ can be analytically continued to a complex manifold $M_\C$.   This means, in particular, that $t$ can be continued to complex values,
keeping the space coordinates real.   
$M_\C$ is only guaranteed to exist in a small neighborhood of $M$, beyond which all sorts of singularities may develop.   Therefore, when we continue $t$ to complex values, we are 
only guaranteed holomorphy if $\Im\,t$ is sufficiently small.  To avoid encountering possible singularities of $M_\C$, we restrict the complexification of $M$ to\footnote{Here in the case of a spatially closed universe, we can assume $\epsilon$
to be a positive constant, but in an open universe, in general it might be necessary to take $\epsilon$ to be a positive function on $S$.   Since $t$ could be multiplied by any $\varrho$-invariant
positive 
real analytic function on $M$, for $\epsilon$ to be constant or non-constant is not really a natural condition in the absence of further information about the function $t$.}
 $|\Im\, t| \leq \epsilon$ for some small $\epsilon>0$.
 
We can now define a Euclidean signature manifold $M_E$ by taking $t=\i t_E$ to be imaginary, while keeping the spatial coordinates of $S$ real.  $M_E$ can be described
more intrinsically as the fixed point set of $\lambda\varrho$, where $\varrho$ is the same time-reversal symmetry as before (analytically continued to act holomorphically on $M_\C$)
and $\lambda$ is complex conjugation of the coordinates (in other words, $\lambda$ is the antiholomorphic symmetry of $M_\C$ that has $M$ as a fixed point set).   In what follows, 
we restrict $M_E$ to the portion with $0\geq t_E\geq -\epsilon$, which in particular serves to avoid the singularities of $M_\C$.
 
      \begin{figure}
 \begin{center}
   \includegraphics[width=2.5in]{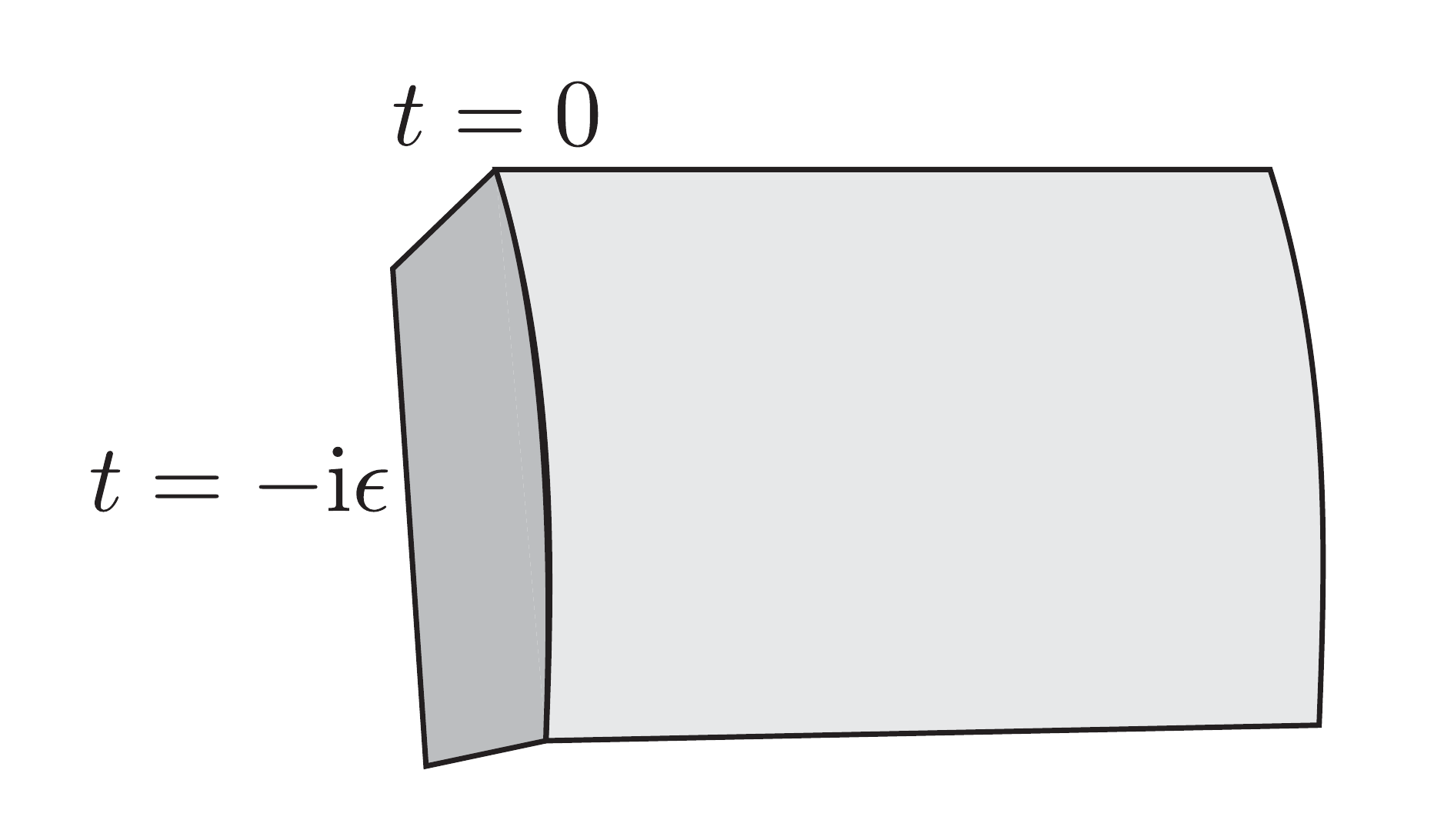}
 \end{center}
\caption{\footnotesize  The lightly shaded region is the $t>0$ portion of a real analytic spacetime of Lorentz signature.   From $t=0$,  we  continue the spacetime
in the direction of imaginary time, truncating it at $t=-\i\epsilon$.   The  Euclidean part of the picture is more darkly shaded.    \label{four}}
\end{figure} 
 
 The idea is now that Euclidean propagation on $M_E$ from  $t_E=-\epsilon$ to $t_E=0$ will map any state at all at $t_E=-\epsilon$ to an analytic state at $t_E=0$.  
An analogous statement in ordinary quantum mechanics is that, if the Hamiltonian $H$ is real analytic (meaning that it maps real analytic wavefunctions to
 real analytic wavefunctions) and  bounded below, then acting with $\exp(-\epsilon H)$, $\epsilon>0$,
 will map any state to a real analytic one.  (For example, if $H$ is the Laplacian on $\R^n$, then $\exp(-\epsilon H)$ maps a delta function to the heat kernel,
 which is a real analytic Gaussian function on $\R^n$.)   Since in general we do not assume $M$ to be time-independent, a closer analog in ordinary quantum mechanics
 would be imaginary time propagation via a  time-dependent Hamiltonian $H(t_E)$, still assumed to be real analytic and bounded below.  Propagation in imaginary time from $t_E=-\epsilon$ to
 $t_E=0$  with a time-dependent Hamiltonian can  be described by a path-ordered product
 $P\exp(-\int_{-\epsilon}^0 \d t_E H(t_E))$, still mapping any state to a real analytic state.
 
In the case of quantum field theory in curved spacetime, since the real time and imaginary time spaces $M$ and $M_E$ meet at $t=t_E=0$, after propagation
through imaginary time from $t_E=-\epsilon$ to $t_E=0$, the state can be continued onto $M$ and interpreted as a state
 in the original Lorentz signature spacetime.   Any initial state at $t_E=-\epsilon $ is expected to 
propagate to a real analytic state on $M$.
One method to define the initial condition in the Euclidean evolution is to impose a local boundary condition at $t_E=-\epsilon$.     In \cite{GW},
in free scalar field theory, the state at $t_E=-\epsilon$ was defined by Dirichlet or Neumann boundary conditions, and it was proved that the resulting state on $M$ is analytic.

Heuristically, one would expect the same construction, with a suitable choice of local boundary condition at $t_E=-\epsilon$,  to work in any physically realistic quantum field theory.
For example, for gauge fields, one can again assume Dirichlet or Neumann boundary conditions at $t_E=-\epsilon$.   At first, one might think that this construction would not apply in a theory such as the Standard Model of particle physics that has chiral fermions, since there is no gauge-invariant local boundary condition
for the fermions of such a theory.  However, if one assumes Dirichlet boundary conditions for the gauge fields, then there is no reason to require the fermion boundary
condition  to be gauge-invariant. So in fact this construction can apply perfectly well in the Standard Model.
 
 What about a spacetime $M$ that does not have a time-reflection symmetry?
  In such a spacetime, one can still pick a real analytic coordinate $t$ such
 that $t=0$ defines a Cauchy hypersurface $S$ and near $S$,
 the metric takes the form
 \be\label{tolmo}\d s^2=-\d t^2+\sum_{i,j=1}^d g_{ij}(\vec x,t)\d x^i \d x^j,\ee
 with a time-dependent spatial metric $g_{ij}(\vec x,t)$.   
  And one can still analytically continue $t$ to complex values.  The difference from the previous case is that the manifold
 $M_E$ defined by keeping $\vec x$  real and taking $t=\i t_E$ to be imaginary no longer has a real  Euclidean signature metric.   However, for small $t_E$,
 the metric on $M_E$ is almost real, and its real part is positive-definite.   A consequence is that the same construction as before is applicable for sufficiently small $\epsilon$, as shown rigorously for bosonic  
 free field theory in \cite{GW}.
 A prototype of this situation in ordinary quantum mechanics, again assuming that $H$ is real analytic and bounded below, is that propagation by $\exp(-(\epsilon\pm \i \epsilon^2)H)$
 will map any state to a real analytic one.  Again, heuristically, one expects that this construction applies equally to non-free theories.   
 
 For free theories, the analytic states obtained this way can be characterized as being annihilated by a maximal commuting subalgebra of the field algebra (analogous to
 a full set of annihilation operators in Minkowski space), and therefore are cyclic for the field algebra.   One expects the same for non-free theories, if only because a generic
 state is cyclic.

\vskip1cm
 \noindent {\it {Acknowledgements}}  Research of EW supported in part by NSF Grant PHY-2207584.

 \bibliographystyle{unsrt}

\end{document}